\documentstyle[aps,12pt]{revtex}
\def\bfig{\begin{figure}}
\def\efig{\end{figure}}
\def\bea{\begin{eqnarray}}
\def\beann{\begin{eqnarray*}}
\def\eea{\end{eqnarray}}
\def\eeann{\end{eqnarray*}}

\begin{document}
\title{Meson-baryon interaction in the quark-gluon exchange framework}
\author{Dimiter Hadjimichef}
\address{Instituto de F\'{\i}sica e Matem\'{a}tica, UFPel, Caixa Postal 354\\
96010-900, Pelotas, RS.}
\maketitle

\begin{abstract}
We extend the Fock-Tani formalism to include the meson-baryon interaction.
The Fock-Tani formalism is a first principle method which allows the use of
field theoretic methods for treating systems of composite particles. An
application of this general result can be the $K^{+}N$ system which is now
reconized as much more suitable system for the investigation of the short
range nature of the hadronic repulsion.
\end{abstract}

\vspace{1.0cm}In the present mesons and baryons are considered to be the
effective degrees of freedom of QCD at hadronic scales. The interaction is
of finite range, attractive at long and intermediate ranges and repulsive at
short distances. Although meson exchange models describing the $NN$
interaction fit remarkably well the known data, the question regarding the
physical origin of the large short range repulsion between the nucleon still
remains. Many meson-exchange models (Paris\cite{paris}, Nijmegen \cite{nij}
and Bonn \cite{bonn}) present a good description of the low energy data. In
the Bonn potential this repulsion is described in terms of an $\omega $
exchange with a coupling constant $g_{NN\omega }^2$ much larger than
obtained from SU(3) estimates. This high value can indicate the presence of
``exotic'' contributions such as: (a) in the Nijmegen meson exchange model
repulsion arises partly from a pomeron exchange \cite{paris}; (b)
calculations involving the relativistic Gross equation, some repulsion is
generated by pairs of nucleon-antinucleon terms \cite{gross}, (c) quark
exchange mechanisms \cite{quark}. In the quark exchange framework the
repulsion has two origins: (1) the antisymmetry of the 6-quark wavefunction,
(2) the spin-spin interaction between quarks. Although the $NN$ data is very
precise one cannot discriminate between many short range scenarios.

It has been argued \cite{kn} that a more suitable investigation of the short
range part of the hadronic repulsion could be made in the $KN$ system. This
system has baryon number equal to 1 so for low energies {$K^{-}N$ is limited
to two channels (}$\pi \Lambda $ and $\pi \Sigma $). The {$K^{+}N$
interaction is short ranged and relatively weak. The one pion exchange
vertex is zero so there is no }$\pi KK$ vertex{\ and 2}$\pi $ exchanges are
weak. In the constituent quark model the $K^{+}=u\,\bar{s}$ and $K^{-}=\,s\,%
\bar{u}$ so the simple quark exchange mechanism can be applied to the {$%
K^{+}N$ system. }Many groups \cite{kn-many} have studied this system in the
S-wave it is necessary to extend this investigation in order to include
higher partial waves, such as the P-waves.

The method we employ in order to introduce the quark-gluon degrees of
freedom is known as the Fock-Tani formalism \cite{ft}. The composite meson
and baryon possess creation operators which include the internal structure
explicitly

{\ 
\[
M_\alpha ^{\dagger }=\Phi _\alpha ^{\mu \nu }q_\mu ^{\dagger }\overline{q}%
_\nu ^{\dagger }\;\;\;{\;;\;{\;\;{\;\;}}B_\alpha ^{\dagger }=\frac 1{\sqrt{6}%
}\Psi _\alpha ^{\mu _1\mu _2\mu _3}q_{\mu _1}^{\dagger }q_{\mu _2}^{\dagger
}q_{\mu _3}^{\dagger }} 
\]
where quarks and antiquarks obey the following anticommutation relations 
\[
\left\{ q_\mu ,q_\nu \right\} =\left\{ \overline{q}_\mu ,\overline{q}_\nu
\right\} =\left\{ q_\mu ,\overline{q}_\nu \right\} =0\;{\;\;}\;{;}\;\;\;{%
\;\;\left\{ q_\mu ,q_\nu ^{\dagger }\right\} =\delta _{\mu \nu
}\;;\;\;\left\{ \overline{q}_\mu ,\overline{q}_\nu ^{\dagger }\right\}
=\delta _{\mu \nu }.} 
\]
The composite meson and baryon operators satisfy non-canonical
(anti)commutation relations 
\[
\left[ M_\alpha ,M_\beta \right] =0{\;{\;;{\;{\;}}}\left[ M_\alpha ,M_\beta
^{\dagger }\right] =\delta _{\alpha \beta }-D_{\alpha \beta }\;\;\;;\;\;\;{%
\left\{ B_\alpha ,B_\beta \right\} =0\;\;\;\;;{\;\;\;}\left\{ B_\alpha
,B_\beta ^{\dagger }\right\} =\delta _{\alpha \beta }-\Delta _{\alpha \beta }%
}} 
\]
} {with 
\begin{eqnarray*}
&&D_{\alpha \beta }=\Phi _\alpha ^{*\mu \nu }\Phi _\beta ^{\mu \sigma }%
\overline{q}_\sigma ^{\dagger }\overline{q}_\nu +\Phi _\alpha ^{*\mu \nu
}\Phi _\beta ^{\rho \nu }q_\rho ^{\dagger }q_\mu \;\  \\
&&\Delta _{\alpha \beta }={3\Psi _\alpha ^{*\mu _1\mu _2\mu _3}\Psi _\beta
^{\mu _1\mu _2\nu _3}q_{\nu _3}^{\dagger }q_{\mu _3}-\frac 32\Psi _\alpha
^{*\mu _1\mu _2\mu _3}\Psi _\beta ^{\mu _1\nu _2\nu _3}q_{\nu _3}^{\dagger
}q_{\nu _2}^{\dagger }q_{\mu _2}q_{\mu _3}}
\end{eqnarray*}
The Fock-Tani formalism introduces ``ideal particles'' which obey canonical
(anti)commutation relations} 
\[
{\left[ m_\alpha ,m_\beta \right] =0\;;\;\;\left[ m_\alpha ,m_\beta
^{\dagger }\right] =\delta _{\alpha \beta
}\;\;\;\;\;\;\;\;\;;\;\;\;\;\;\;\left\{ b_\alpha ,b_\beta \right\}
=0\;\;;\;\;\left\{ b_\alpha ,b_\beta ^{\dagger }\right\} =\delta _{\alpha
\beta }} 
\]
{\ This way one can transform the composite state into a ideal state by 
\begin{eqnarray*}
U_M(-\frac \pi 2)M_\alpha ^{\dagger }|0\rangle =m_\alpha ^{\dagger
}|0\rangle \;\;\;{\;\;\;{{\;\;\;}\;{\;\;\;}}U_B(-\frac \pi 2)B_\alpha
^{\dagger }|0\rangle =b_\alpha ^{\dagger }|0\rangle }
\end{eqnarray*}
where\ 
\begin{eqnarray*}
U_M(t)=\exp \left[ tF_M\right] {\;{\;}}\;{{{\;{\;}}}};{{{{\;{\;}}}}}\;{{\;{\;%
}}}U_B(t)=\exp \left[ tF_B\right] \;
\end{eqnarray*}
F}$_{M\text{ }}${and F}$_B${\ \ are the generators of the transformation 
\[
F_M=m_\alpha ^{\dagger }O_\alpha ^M-O_\alpha ^{M\dagger }m_\alpha \;\;\;\;{%
\;;\;{\;\;\;}\;F_B=b_\alpha ^{\dagger }O_\alpha ^B-O_\alpha ^{B\dagger
}b_\alpha } 
\]
with 
\begin{eqnarray*}
O_\alpha ^M &=&M_\alpha +\frac 12D_{\alpha \tau }M_\tau -\frac 12M_\omega
^{\dagger }\left[ D_{\omega \tau },M_\alpha \right] M_\tau \  \\
O_\alpha ^B &=&B_\alpha +\frac 12\Delta _{\alpha \tau }B_\tau -\frac 12%
B_\omega ^{\dagger }\left[ \Delta _{\omega \tau },B_\alpha \right] B_\tau .
\end{eqnarray*}
}

{The generator of the meson-baryon transformation is given by} {\ 
\begin{equation}
F=\widetilde{F}_M+\widetilde{F}_B=m_\alpha ^{\dagger }\widetilde{O}_\alpha
^M-\widetilde{O}_\alpha ^{M\dagger }m_\alpha +b_\alpha ^{\dagger }\widetilde{%
O}_\alpha ^B-\widetilde{O}_\alpha ^{B\dagger }b_\alpha   \label{ft-mb}
\end{equation}
where} {\ 
\[
\widetilde{O}_\alpha ^M=O_\alpha ^M+G_\alpha ^M{\;}\;{\;{{{\;{\;}}}\;;\;{\;\;%
{\;}\;\;}}}\;{\widetilde{O}_\alpha ^B=O_\alpha ^B+G_\alpha ^B}
\]
with} {\ 
\begin{eqnarray*}
G_\alpha ^M &=&-\frac 12\ \left[ M_\alpha ,B_\tau ^{\dagger }\right] B_\tau -%
\frac 12B_\omega ^{\dagger }\left[ M_\alpha ,\Delta _{\omega \tau }\right]
B_\tau {+\frac 12\ M_\omega ^{\dagger }\left[ M_\alpha ,\left[ M_\omega
,B_\tau ^{\dagger }\right] \right] B_\tau } \\
&&\ -\frac 12B_\omega ^{\dagger }\left[ B_\omega ,D_{\alpha \tau }\right]
M_\tau  \\
G_\alpha ^B &=&\ \frac 12\left[ M_\tau ^{\dagger },B_\alpha \right] M_\tau -%
\frac 12M_\omega ^{\dagger }\left[ B_\alpha ,D_{\omega \tau }\right] M_\tau {%
-\ \frac 12M_\omega ^{\dagger }\left[ M_\omega ,\Delta _{\tau \alpha
}\right] B_\tau }
\end{eqnarray*}
} {\ In order to obtain the effective meson-baryon potential one has to use (%
\ref{ft-mb}) in a set of Heisenberg-like equations for the basic operators $%
m,b,\widetilde{O}^M,\widetilde{O}^B,q,\overline{q\text{ }}$: } {\ 
\[
\frac{dm_\alpha \left( t\right) }{dt}=\left[ m_\alpha ,F\right] =\widetilde{O%
}_\alpha ^M\;{\;{{{\;\;\;}\;}}}\;;\;\;{{{\;}\;}}\;{\frac{db_\alpha \left(
t\right) }{dt}=\left[ b_\alpha ,F\right] =\widetilde{O}_\alpha ^B}
\]
The equations of motion for $\widetilde{O}^M$ and $\widetilde{{\ }O}^B$ are 
\[
\frac{d\widetilde{O}_\alpha ^M\left( t\right) }{dt}=\left[ \widetilde{O}%
_\alpha ^M,F\right] \ =-m_\alpha {{{{{\;\;\;\;}\;}}}\;;\;\;\;{{{\;}\;}}\frac{%
d\widetilde{O}_\alpha ^B\left( t\right) }{dt}=\left[ \widetilde{O}_\alpha
^B,F\right] \ =-b_\alpha }
\]
The simplicity of these equations are not present in the equations for $q$
and $\overline{q}$: } {\ 
\[
\frac{dq_\mu \left( t\right) }{dt}=\left[ q_\mu ,F\right] \ ={\dot{q}_\mu
^M\left( t\right) +\dot{q}_\mu ^B\left( t\right) +\left[ q_\mu ,m_\beta
^{\dagger }G_\beta ^M-G_\beta ^{M\dagger }m_\beta \right] +{\left[ q_\mu
,b_\beta ^{\dagger }G_\beta ^B-G_\beta ^{B\dagger }b_\beta \right] }}
\]
and} {\ 
\[
\frac{d\overline{q}_\mu \left( t\right) }{dt}=\left[ \overline{q}_\mu
,F\right] =\ {\dot{\overline{q}}_\mu ^M\left( t\right) +\dot{\overline{q}}%
_\mu ^B\left( t\right) \ +\left[ \overline{q}_\mu ,m_\beta ^{\dagger
}G_\beta ^M-G_\beta ^{M\dagger }m_\beta \right] +{\left[ \overline{q}_\mu
,b_\beta ^{\dagger }G_\beta ^B-G_\beta ^{B\dagger }b_\beta \right] }}
\]
The solutions for these equations can be found order by order in the
wavefunctions. So for zero order one has ${{q}_\mu ^{(0)}=q_\mu }$} , {${{%
\overline{q}_\nu ^{(0)}=}{{\overline{q}_\nu }}}$} {\ 
\[
m_\alpha ^{(0)}(t)=M_\alpha \;{\rm sen\,}t+m_\alpha \;\cos t\;{{{\;;}\;}%
b_\alpha ^{(0)}(t)=B_\alpha \;{\rm sen\,}t+b_\alpha \;\cos t\;{.}}
\]
} In first order {$m_\alpha ^{(1)}{=0,}$ ${b_\alpha ^{(1)}=0}$ and} {\ 
\begin{eqnarray*}
q_\mu ^{(1)}(t) &=&-\sqrt{\frac 32}\Psi _\alpha ^{\mu \mu _2\mu _3}q_{\mu
_2}^{\dagger }q_{\mu _3}^{\dagger }\left[ \ b_\alpha \;{\rm sen\,}t+B_\alpha
\;(1-\cos t)\right] {-\Phi _\alpha ^{\mu \nu }\bar{q}_\nu ^{\dag }\left[ \
m_\alpha \;{\rm sen\,}t+M_\alpha \;(1-\cos t)\right] } \\
{\overline{q}_\nu ^{(1)}(t)} &=&{\Phi _\alpha ^{\mu _1\nu }q_{\mu _1}^{\dag
}\left[ \ m_\alpha \;{\rm sen\,}t+M_\alpha \;(1-\cos t)\right] .}
\end{eqnarray*}
} {\ In second order we find } 
\begin{eqnarray*}
q_\mu ^{(2)}(t) &=&\sqrt{6}\,\Psi _\alpha ^{\mu \mu _2\mu _3}\Phi _\beta
^{*\mu _2\nu }{\left[ M_\beta ^{\dagger }\,q_{\mu _3}^{\dagger }\bar{q}_\nu
B_\alpha \,(\frac 12\cos ^2t-\cos t+\frac 12)\right. } \\
&&+M_\beta ^{\dagger }\,q_{\mu _3}^{\dagger }\bar{q}_\nu b_\alpha \,(-\frac 1%
2{\rm sen\,}t\cos t\ +{\rm sen\,}t)+\frac 12m_\beta ^{\dagger }\,q_{\mu
_3}^{\dagger }\bar{q}_\nu b_\alpha \,{\rm sen}^2{\rm \,}t \\
&&\left. +\frac 12m_\beta ^{\dagger }\,q_{\mu _3}^{\dagger }\bar{q}_\nu
B_\alpha \,{\rm sen\,}t\cos t\right] +\left( \Psi ^{*}\Psi +\Phi ^{*}\Phi
\right)  \\
\overline{q}_\nu ^{(2)}(t) &=&\sqrt{\frac 32}\,\Phi _\alpha ^{\ \mu _1\nu
}\Psi _\beta ^{*\mu _1\mu _2\mu _3}{\left[ B_\beta ^{\dagger }\,q_{\mu
_3}^{\ }q_{\mu _2}M_\alpha \,(-\frac 12\cos ^2t+\cos t-\frac 12)\right. } \\
&&{+\ B_\beta ^{\dagger }\,q_{\mu _3}^{\ }q_{\mu _2}m_\alpha \,(\frac 12{\rm %
sen\,}t\cos t\ -{\rm sen\,}t)-\frac 12\ b_\beta ^{\dagger }\,q_{\mu _3}^{\
}q_{\mu _2}m_\alpha {\rm sen}^2{\rm \,}t} \\
&&{\left. +\frac 12\ \,b_\beta ^{\dagger }\,q_{\mu _3}^{\ }q_{\mu
_2}M_\alpha {\rm sen\,}t\cos t\right] }+\left( \Psi ^{*}\Psi +\Phi ^{*}\Phi
\right) 
\end{eqnarray*}
We shall omit the third order operators, but they can be calculated in the
same way. {\ The meson-baryon potential can be obtained applying in a
standard way the Fock-Tani transformed operators to the microscopic
Hamiltonian \cite{ft} 
\begin{eqnarray*}
H &=&T(\mu )\,q_\mu ^{\dagger }\,q_\mu +T(\nu )\,\overline{q}_\nu ^{\dagger
}\,\overline{q}_\nu {+\frac 12V_{qq}(\mu \nu ;\sigma \rho )\,q_\mu ^{\dagger
}\,q_\nu ^{\dagger }q_\rho q_\sigma } \\
&&+\frac 12V_{\overline{q}\overline{q}}(\mu \nu ;\sigma \rho )\,\overline{q}%
_\mu ^{\dagger }\,\overline{q}_\nu ^{\dagger }\overline{q}_\rho \overline{q}%
_\sigma {+\ V_{q\overline{q}}(\mu \nu ;\sigma \rho )q_\mu ^{\dagger }\,%
\overline{q}_\nu ^{\dagger }\overline{q}_\rho q_\sigma }
\end{eqnarray*}
where one finds 
\[
V_{{\rm meson-baryon}}=\sum_{i=1}^4\;V_i(\alpha \beta ;\delta \gamma )\;{{%
m_\alpha ^{\dagger }\,}b_\beta ^{\dagger }\,m_\gamma \,b_\delta }
\]
and} {\ 
\begin{eqnarray*}
V_1{(\alpha \beta ;\delta \gamma )} &=&-3V_{qq}(\mu \nu ;\sigma \rho )\,\Phi
_\alpha ^{*\mu \nu _2}\Psi _\beta ^{*\nu \mu _2\mu _3}\Phi _\gamma ^{\rho
\nu _2}\Psi _\delta ^{\sigma \mu _2\mu _3}\;\  \\
V{_2{(\alpha \beta ;\delta \gamma )}} &=&{-3V_{q\overline{q}}(\mu \nu
;\sigma \rho )\,\Phi _\alpha ^{*\mu _1\nu }\Psi _\beta ^{*\mu \mu _2\mu
_3}\Phi _\gamma ^{\sigma \rho }\Psi _\delta ^{\mu _1\mu _2\mu _3}} \\
V{_3{(\alpha \beta ;\delta \gamma )}}\text{ } &=&{-3V_{qq}(\mu \nu ;\sigma
\rho )\,\Phi _\alpha ^{*\mu \nu _2}\Psi _\beta ^{*\mu _1\nu \mu _3}\Phi
_\gamma ^{\mu _1\nu _2}\Psi _\delta ^{\sigma \rho \mu _3}} \\
V{_4{(\alpha \beta ;\delta \gamma )}}\text{ } &=&{-6V_{q\overline{q}}(\mu
\nu ;\sigma \rho )\,\Phi _\alpha ^{*\nu _1\nu }\Psi _\beta ^{*\mu _1\mu \mu
_3}\Phi _\gamma ^{\mu _1\rho }\Psi _\delta ^{\nu _1\sigma \mu _3}.}
\end{eqnarray*}
}

\underline{{\ Future perspectives}}

Using the Fock-Tani formalism we have derived an effective meson-baryon
potential. This potential can now be used to calculate the interesting $%
K^{+}N$ system, which is more suitable for the investigation of the short
range part of the hadronic repulsion.

\underline{{\bf Acknowledgements}}

The author was partially supported by FAPERGS and FAPESP (Proc. N.
1998/2249-4).

\newpage\


\begin{references}
\bibitem{paris}  M. Lacombe, B. Loiseau, J. M. Richard, R. Vinh Mau, J.
C\^{o}t\'{e}, P. Pir\`{e}s and R. Toureil, {\sl Phys. Rev}{\it .} {\bf C21},
861 (1980).

\bibitem{nij}  M. M. Nigels, T. A. Rijken and J. J. deSwart, {\sl Phys. Rev}%
{\it .} {\bf D17}, 768 (1978).

\bibitem{bonn}  R. Machleidt, K. Holinde, Ch. Elster, {\sl Phys. Rep}{\it .} 
{\bf 149}, 1 (1987).

\bibitem{gross}  F. Gross, {\sl Phys. Rev}{\it .} {\bf 186}, 1448 (1969).

\bibitem{quark}  M. Oka and K. Yazaki, {\sl Prog. Theor. Phys}. {\bf 66},
556 (1981); {\bf 66}, 572 (1981); A. Faessler, F. Fernandes, G. L\"{u}beck
and K. Shimizu, {\sl Phys. Lett.} {\bf B112}, 201 (1982).

\bibitem{kn}  R. B\"{u}ttgen, K. Holinde, D. Lohse, A. M\"{u}ller-Groeling,
J. Speth, P. Wyborny, {\sl Z. Phys}. {\bf C46} - Particles and Fields, S167
(1990).

\bibitem{kn-many}  J. Bender, H. G. Dosch, H. J. Pirner and H. G. Kruse, 
{\sl Nucl. Phys.} {\bf A414}, 359 (1984), D. Mukhopadyay and H. J. Pirner, 
{\sl Nucl. Phys.} {\bf A442}, 605 (1985), R. K. Campbell and D. Robson, {\sl %
Phys. Rev}{\it .} {\bf D36}, 2682 (1987), T. Barnes and E. S. Swanson, {\sl %
Phys. Rev}{\it .} {\bf C49}, 1166 (1994).

\bibitem{ft}  D. Hadjimichef, G. Krein, S. Szpigel, J. S. da Veiga, {\sl %
Phys. Lett. }{\bf B367},{\sl \ }317 (1996), M. D. Girardeau, G. Krein, D.
Hadjimichef,{\sl \ Mod. Phys. Lett.} {\bf A11}, 1121 (1996), D. Hadjimichef,
G. Krein, S. Szpigel, J. S. da Veiga, {\sl Ann. of Phys}. {\bf 268}, number
1, 105 (1998).
\end{references}
\end{document}